\documentclass{iaus}
\usepackage{graphicx}

\title[Star Formation in massive LSBGs] 
{Star Formation in Massive Low Surface Brightness Galaxies}

\author[O'Neil]   
{K. O'Neil $^1$}

\affiliation{$^1$ NRAO, PO Box 2, Green Bank, WV 24944, USA \break email: koneil@nrao.edu }

\pubyear{2007}
\volume{244}  
\jname{Dark Galaxies and Lost Baryons IAU Symposium}
\editors{Davies, et al eds.}
\begin{document}

\maketitle

\begin{abstract}
Massive low surface brightness galaxies have disk central surface brightnesses at least one magnitude fainter
than the night sky, but total magnitudes and masses that show they are among the largest galaxies known.
Like all low surface brightness (LSB) galaxies, massive LSB galaxies are often in the midst of star formation
yet their stellar light has remained diffuse, raising the question of how star formation is proceeding
within these galaxies.  We have undertaken a multi-wavelength study to clarify the structural parameters
 and stellar and gas content of these enigmatic systems.  The results of these studies, which include HI,
CO, optical, near UV, and far UV images of the galaxies will provide the most in depth study 
done to date of how, when, and where star formation proceeds within this unique subset of the galaxy population.
\end{abstract}

\firstsection 
\section{Introduction}\label{sec:intro}

Low surface brightness galaxies are typically defined as galaxies with observed disk central surface brightness of $\mu_B(0) \ge $ 23.0 mag arcsec$^{-2}$, more than 3$\sigma$ from the \cite{freeman70} value of 21.65 $\pm$ 0.65.  For the purpose of this paper, a {\it massive} low surface brightness galaxy will be defined as a low surface brightness galaxies with M$_B$ $<$ $-$19 and/or M$_{HI}$ $>$ 10$^9$ M$_\odot$.  While this definition is not particularly strict, it ensures that none of the galaxies studied are dwarf systems and that any truly massive galaxies (e.g. systems with M$_{HI}$ $>$ 10$^{10.2}$ M$_\odot$) are included within the sample.

The first massive low surface brightness (LSB) galaxy reported was Malin 1 (\cite[Bothun, \etal\ 1987]{bothun87}).  Since the discovery of Malin 1, a number of other massive LSB galaxies have been found (e.g. \cite[Davies, Phillipps, \& Disney 1988]{davies88}; \cite[Bothun, \etal\ 1990]{bothun90}; \cite[Sprayberry, \etal\ 1993]{sprayberry93}), culminating in a catalog published in 1995 describing the eight known massive LSB galaxies (\cite[Sprayberry, \etal\ 1995]{sprayberry95}).  In the subsequent 10 years a handful of additional massive LSB galaxies have been identified (e.g. \cite[Walsh, Stavely-Smith, \& Osterloo 1997]{walsh97}), but the total number of such systems has been low, making any broad study of the systems difficult.

\begin{figure}
\centering
\resizebox{6.5cm}{!}{\includegraphics{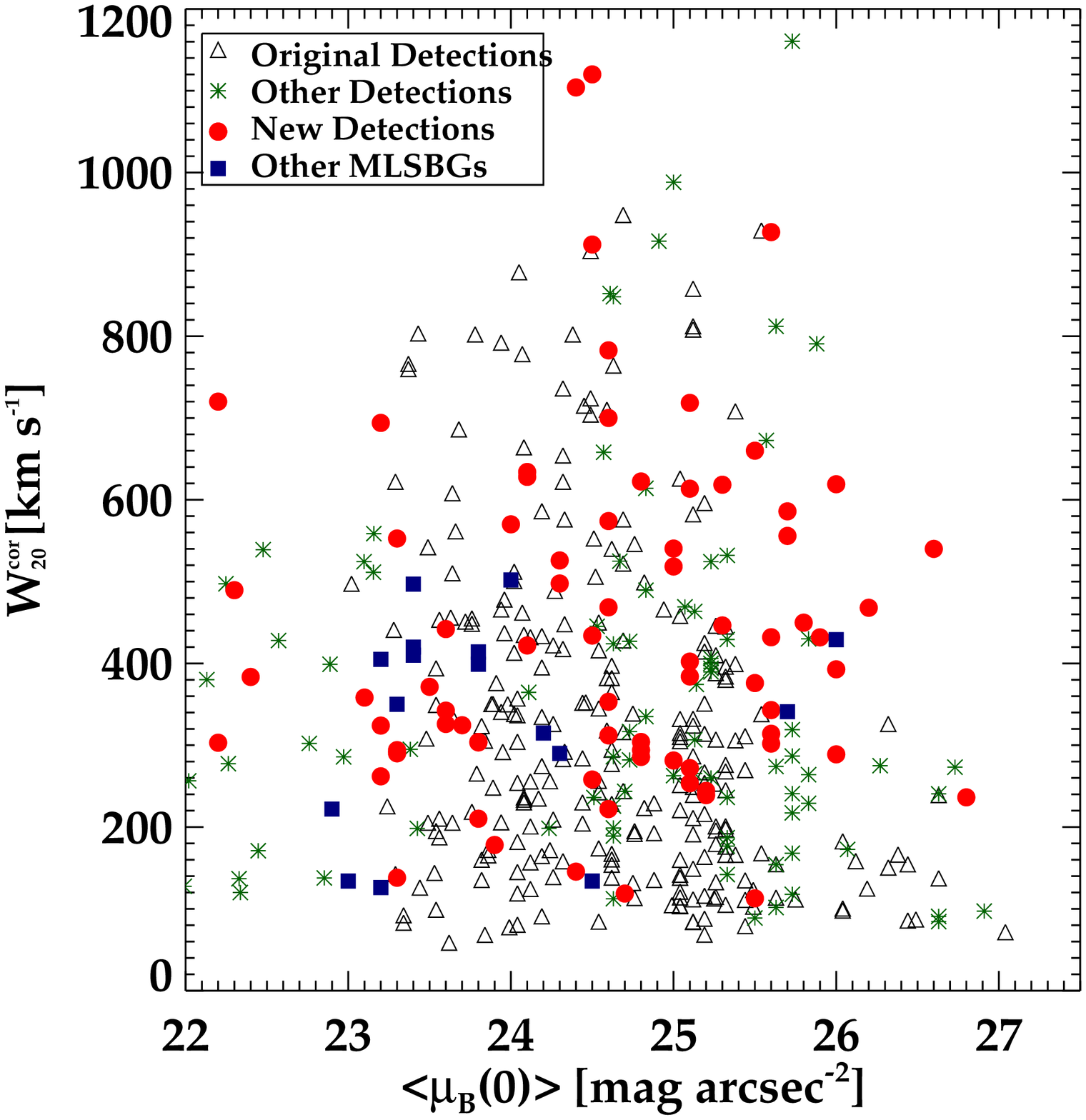} }
\resizebox{6.5cm}{!}{\includegraphics{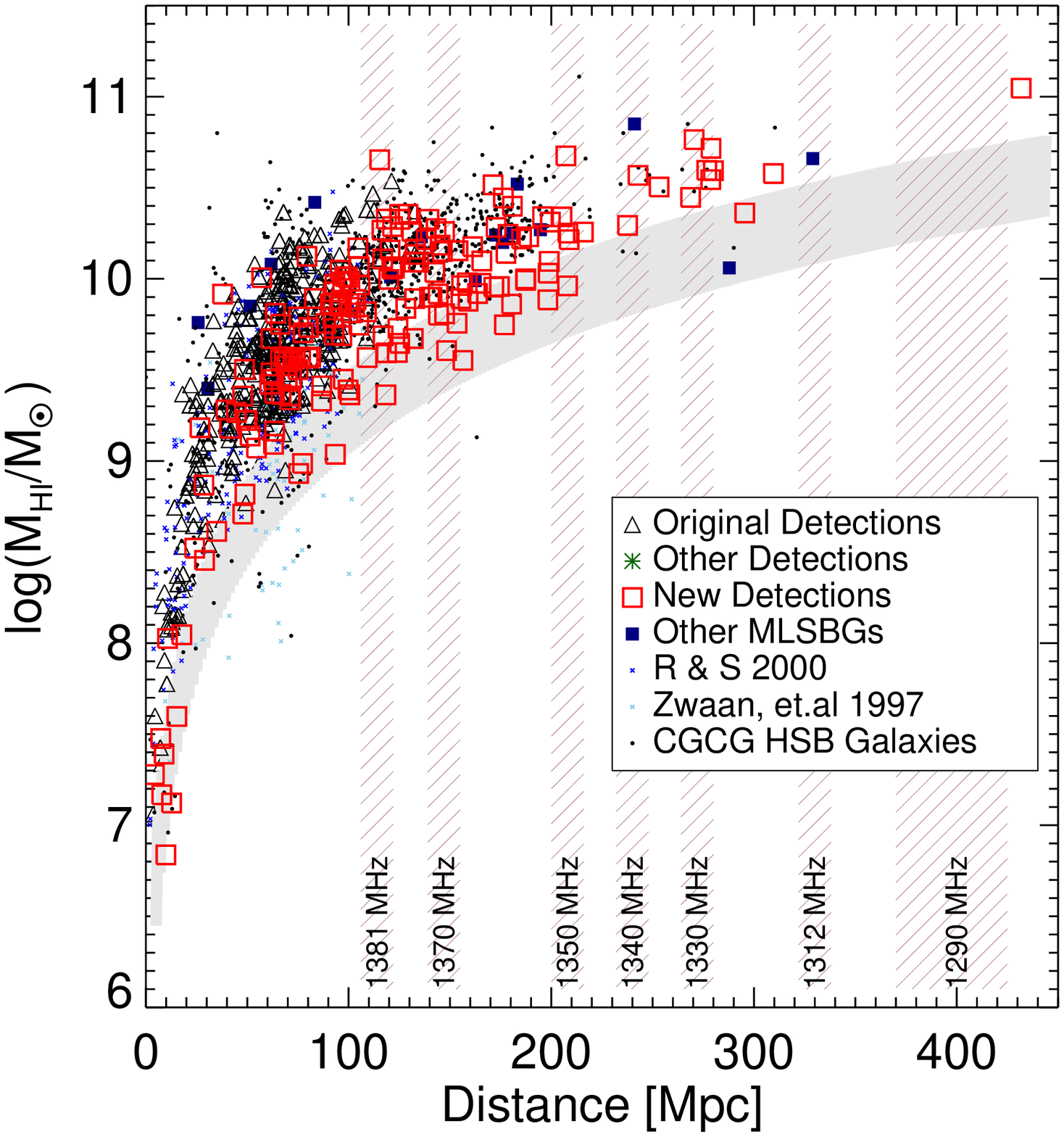} }
\caption[]{{\it Left:} Surface brightness versus (inclination corrected) velocity widths both for all the galaxies in the \cite{oneil04} LSB sample (red circles) and for the 16 massive LSB galaxies with previously cataloged HI properties (blue squares -- \cite[Matthews, van Driel, \& Monnier-Ragaigne 2001]{matthews01b}; \cite[Sprayberry, \etal\ 1995]{sprayberry95}).  The black triangles and green crosses are other LSB galaxies in the catalog of \cite{bothun85} from which the \cite{oneil04} sample was taken.
The inclination correction applied is simply
W$_{20}^{corr}$ = W$_{20}$/sin({\it i}).  To avoid overcorrection, any inclination less than 30$^\circ$ has been set to 30$^\circ$ for the purpose of this correction. Note that the extremely high values of W$_{20}^{corr}$ may be due to an underestimate of the galaxy's inclination. 
{\it Right:} Distance (${velocity}\over{H_0}$) plotted against total HI mass for all the galaxies in \cite{oneil04} (red open squares). For comparison, objects found in the two Arecibo blind HI surveys (\cite[Rosenberg \& Schneider 2000]{rosenberg00}; \cite[Zwaan, \etal\ 1997]{zwaan97}) and  the \cite{giov93} CGCG survey are also plotted.  The gray line on the plot indicates the approximate survey limit for the O'Neil, \etal\ observations, where the line thickness represents the varying sensitivities and velocity widths found herein.  Finally, the diagonal lines show the primary regions affected by radio frequency interference (RFI).  The 16 other massive LSB galaxies with published HI masses (from \cite[Matthews, van Driel, \& Monnier-Ragaigne 2001]{matthews01b} and \cite[Sprayberry, \etal\ 1995]{sprayberry95}) are also plotted (filled blue squares).  The black triangles and green crosses are other LSB galaxies in the catalog of \cite{bothun85} from which the \cite{oneil04} sample was taken.
}\label{fig:oneil04}
\end{figure}

Over the past few years, though, the total number of massive LSB galaxies known has increased significantly.  \cite{oneil04} recently published a catalog of LSB galaxies which had previously been undetected in HI.  Of the more than 100 galaxies observed, 38 fall into the category of massive LSB galaxies, a result which more than tripled the total number known (Figure~\ref{fig:oneil04}).  The catalog of massive LSB galaxies is continuing to grow as more surveys are done to look for these systems. \cite{oneil08} are completing an HI survey of more than 300 massive LSB galaxies with unknown redshifts using the Arecibo, Nan\c{c}ay and Green Bank single dish radio telescopes.  The survey has resulted in more than 190 detections, out of which at least 170 have M$_{HI}$ $>$ 10$^9$ M$_\odot$, and a few have M$_{HI}$ $>$ 10$^{10.2}$ M$_\odot$.

As a result of the recent studies, it is clear that while massive LSB galaxies do not dominate the galaxy counts in the nearby Universe, there are clearly enough of these intriguing systems to make understanding their formation and evolution an important component in all galaxy formation and evolution theories.  Additionally, the catalog of known massive LSB galaxies is finally large enough to make a broad study of these systems feasible.

\section{Dark Matter and Massive LSB Galaxies}

Massive LSB galaxies are not optically dark, as evidenced by their high integrated 
light found (M$_B < -19$).  Yet the study of dark matter in the Universe
necessitates the study of massive LSB galaxies for the same reason the study of less massive
LSB systems is necessary -- massive LSB galaxies, like their less massive counterparts, appear to
be dominated by dark matter, even in their innermost radii.  In the most comprehensive study
done to date, \cite[Pickering, \etal\ (1997, 1998)]{pickering97} show that massive LSB galaxies must have a high dark matter content to match the observed (HI and H$\alpha$) rotation curves. 
The only other option is to let the stellar mass-to-luminosity ratio rise to 10-20
M$_\odot$/L$_\odot$, which results in considerably poorer fits.  Massive LSB galaxies, then,
are not dark galaxies, and they do not solve the ''missing matter'' problem (although they
do contribute to the missing matter solution -- \cite[see O'Neil, \etal\ 2004 ]{oneil04})
but they do offer extreme environments in which dark matter and star formation theories can be tested.

\section{Global Properties of Massive LSB Galaxies}

\subsection{Distribution \& Environment}
The distribution of LSB galaxies follows the same large scale pattern as all other galaxies.  That is, LSB galaxies do not fill in the galaxy voids found in the large scale structure of the Universe.  LSB galaxies do, however, tend to lie on the outer edges of the galaxy distribution (\cite[Rosenbaum \& Bomans 2004]{rosenbaum04}).  On smaller scales ($<$ 2-5 Mpc), though, LSB galaxies are more isolated than high surface brightness (HSB) galaxies, based off nearest neighbor analysis (\cite[Rosenbaum \& Bomans 2004]{rosenbaum04}; \cite[Bothun, \etal\ 1993]{bothun93}; See also paper by Bomans in this volume).   

\subsection{Optical morphology}
\begin{figure}
\centering
\resizebox{6.5cm}{!}{\includegraphics{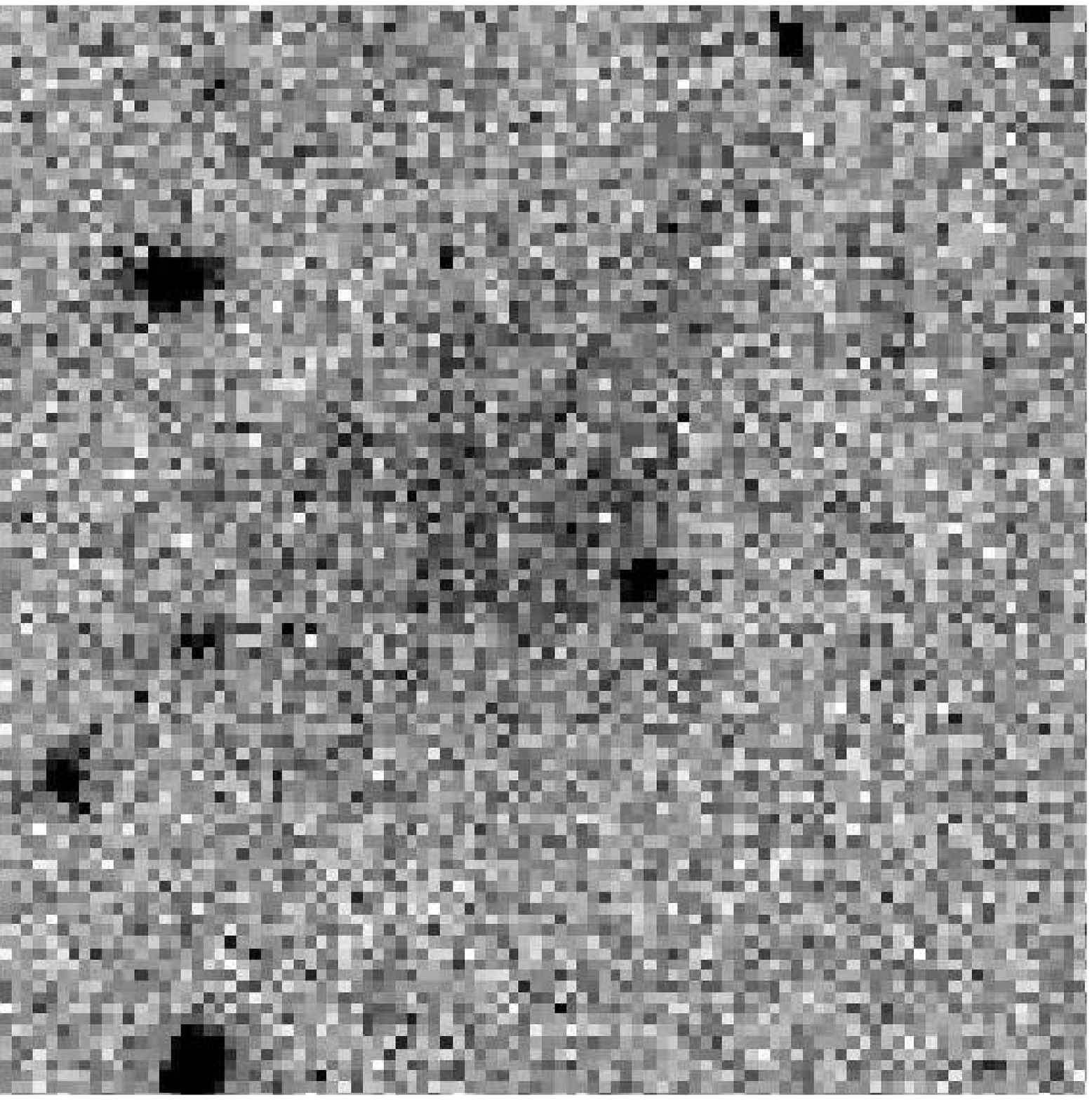} }
\resizebox{6.5cm}{!}{\includegraphics{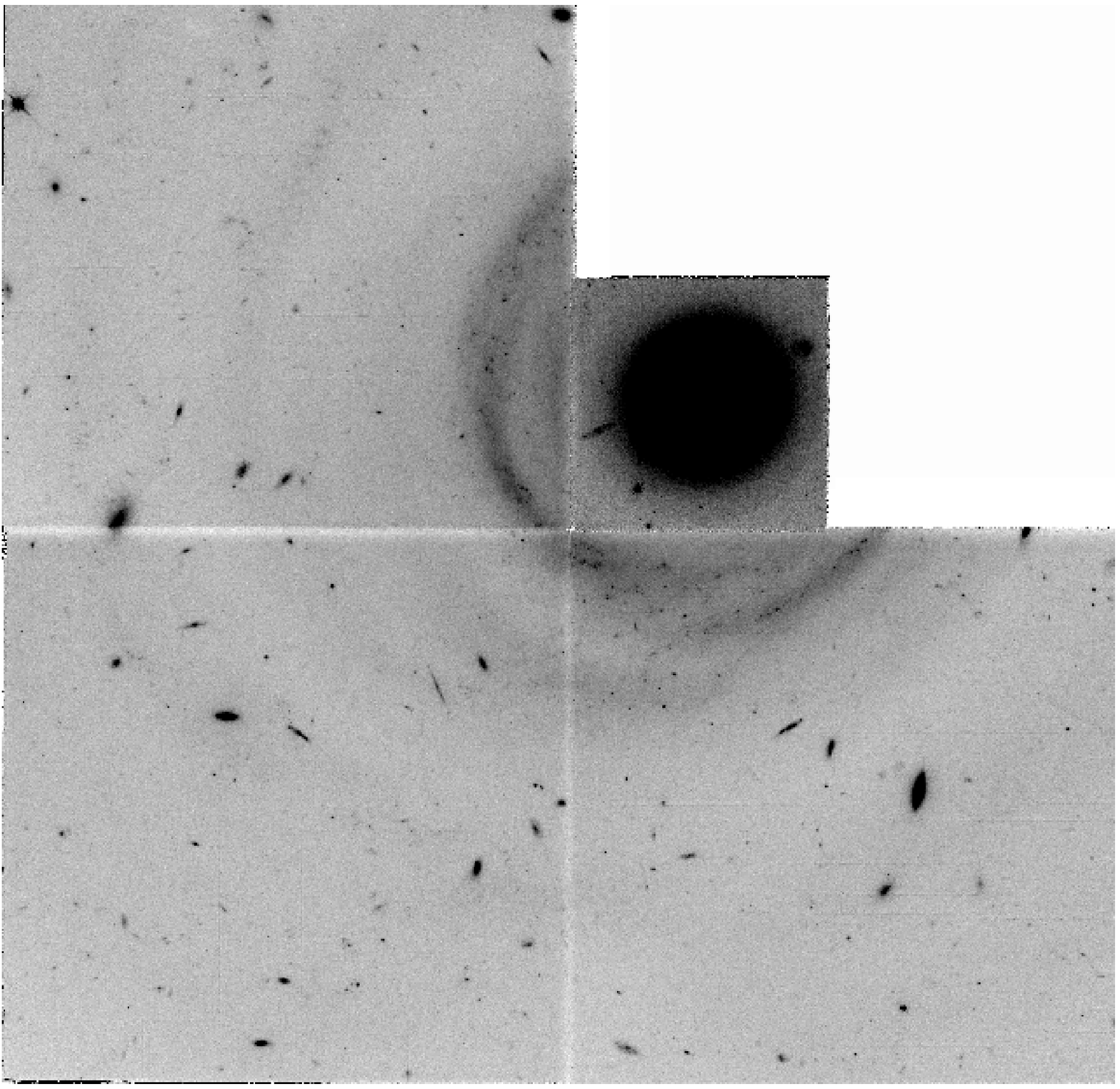} }
\caption[]{Optical images of two low surface brightness galaxies.  On left is $\left[{OBC97}\right]$P02-4, a galaxy with $\mu_B(0)$ = 25.1 mag arcsec$^{-2}$ and m$_B$=17.4 (velocity is unknown -– from \cite[O'Neil, Bothun, \& Cornell 1997]{oneil97}).  
At right is an HST F814W image of UGC 06614, a galaxy with $\mu_B$=24.3 mag arcsec$^{-2}$ and M$_B=-$19.7 (\cite[Mehta \& O'Neil 2006]{mehta06}; \cite[McGaugh \& Bothun 1994]{mcgaugh94}).
}\label{fig:morphology}
\end{figure}

Morphologically, massive LSB galaxies have bright central bulges with diffuse disks which typically contain distinct spiral arms.  This is in contrast to many of the less massive LSB galaxies which often lack any discernible nuclear bulge and for which deep images often cannot distinguish spiral arms (Figure~\ref{fig:morphology}).  Note that no massive LSB galaxy has been found to date without a distinct central bulge.  This may be a result of selection effects, as all massive LSB galaxies were first found in the optical, or it may be due to the large gravitational potential at the center of the massive systems.

\subsection{HI Content}
\begin{figure}
\centering
\resizebox{6.5cm}{!}{\includegraphics{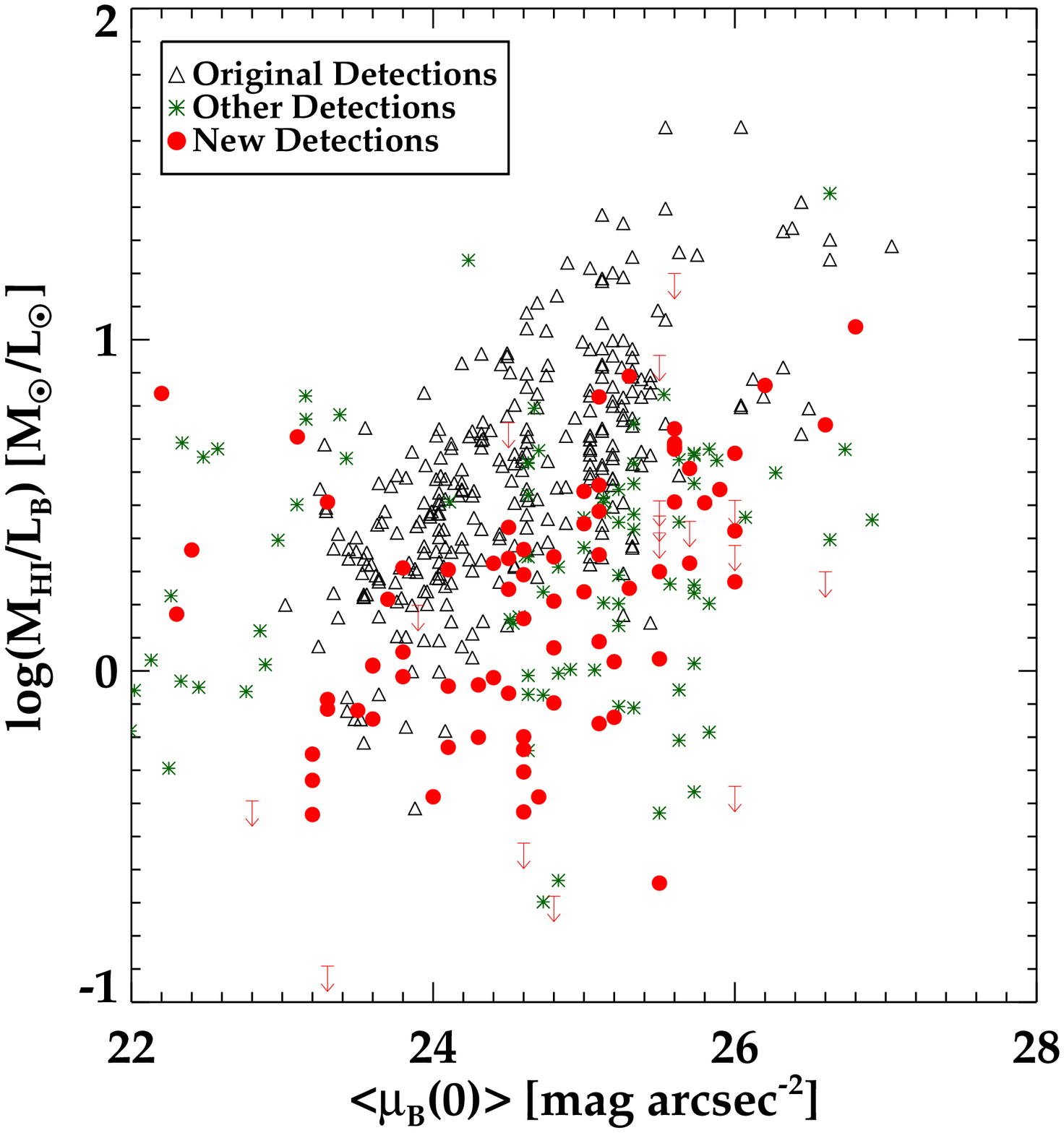} }
\resizebox{6.5cm}{!}{\includegraphics{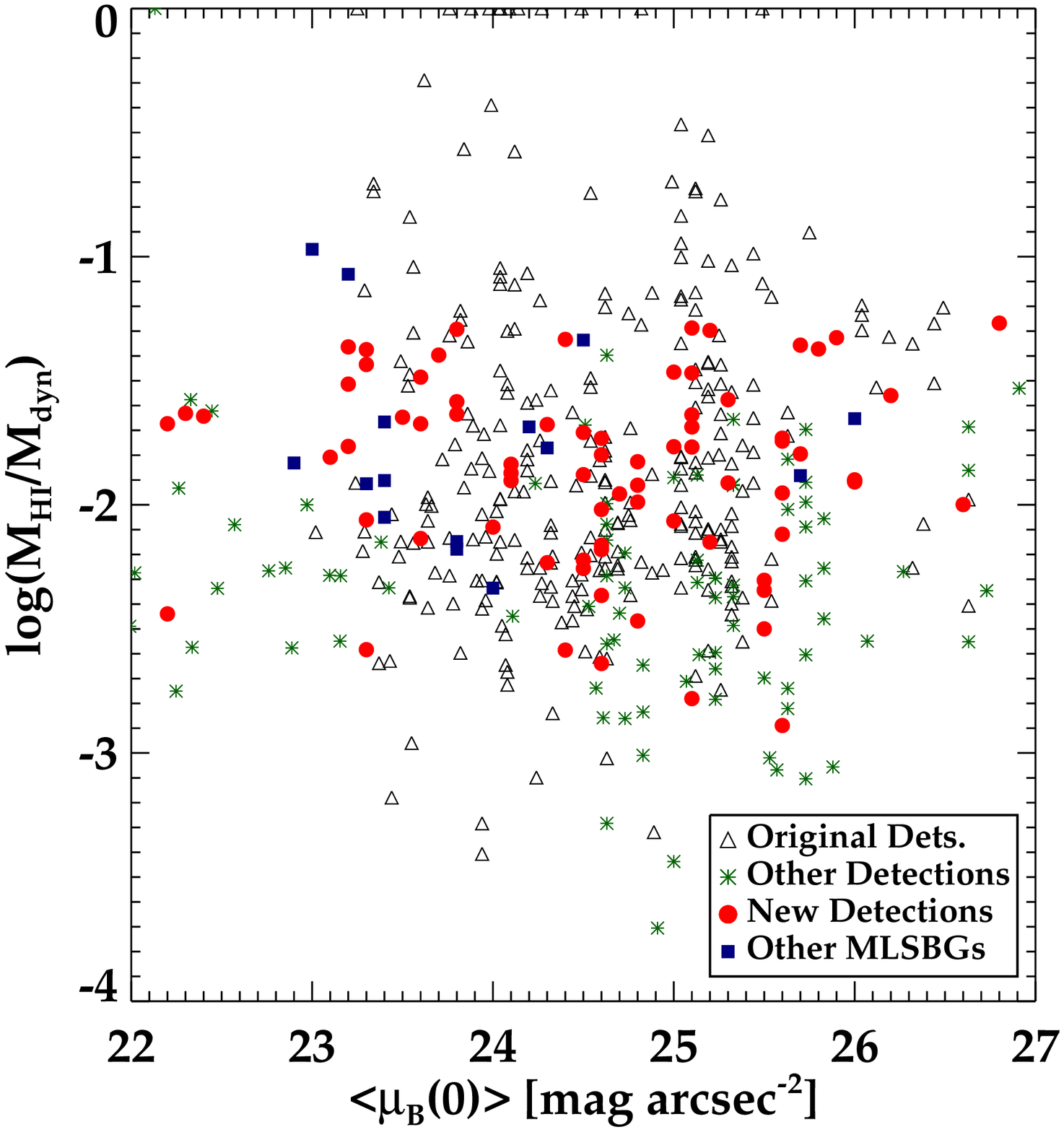} }
\caption[]{Surface brightness versus M$_{HI}$/L$_B$ (left) and M$_{HI}$/M$_{dyn}$ (right) 
for all the galaxies in the \cite{oneil04} LSB sample (red circles).
The black triangles and green crosses are other LSB galaxies in the catalog of \cite{bothun85} from which the \cite{oneil04} sample was taken.
}\label{fig:MHI}
\end{figure}

The M$_{HI}$/L$_B$ ratio for galaxies increases with decreasing surface brightness, a trend which holds true regardless of the mass of the galaxy.  However, the M$_{HI}$/M$_{dyn}$ ratio of the galaxies (with M$_{dyn}$ typically estimated from 1.4$\times$D$_{25}$ and W$_{20}$) does not appear to change with changing surface brightness.  These trends can be seen in Figure~\ref{fig:MHI}.  Here the average surface brightness, rather than the central surface brightness, is used, due to lack of information regarding the central surface brightness of many of the galaxies in the sample.  An analysis of using the average surface brightness to define LSB galaxies is given in \cite{oneil07}.  

Like their less massive counterparts, the distribution of HI gas within massive LSB galaxies typically follows the global morphology defined by the galaxies' stars, although the peaks of the HI distribution do not always lie on top the peaks of the stellar distribution.  Additionally, the HI density of massive LSB galaxies is often at or below the Kennicutt-Toomre criteria (\cite[Kennicutt 1983]{kennicutt83} and \cite[Toomre 1964]{toomre64}) for star formation.  \cite{pickering97} shows only a few peaks in the HI density which exceed the Kennicutt-Toomre criteria, while most of the gas in the galaxies studied falls below the 10$^{20}$ cm$^{-2}$ threshold found for star formation in most galaxies, regardless of their surface brightness.

\begin{figure}
\centering{
\resizebox{4.4cm}{!}{\includegraphics{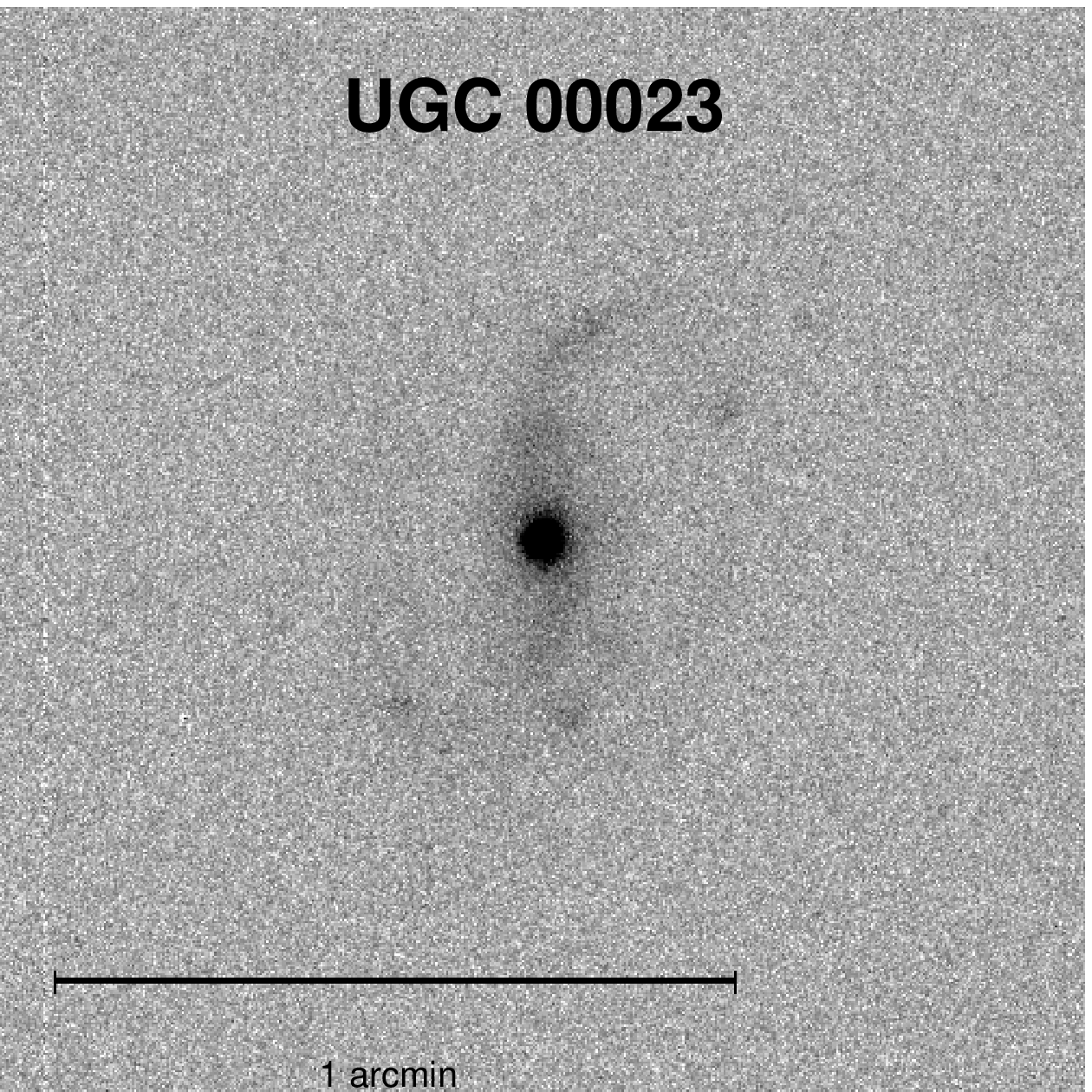} }
\resizebox{4.4cm}{!}{\includegraphics{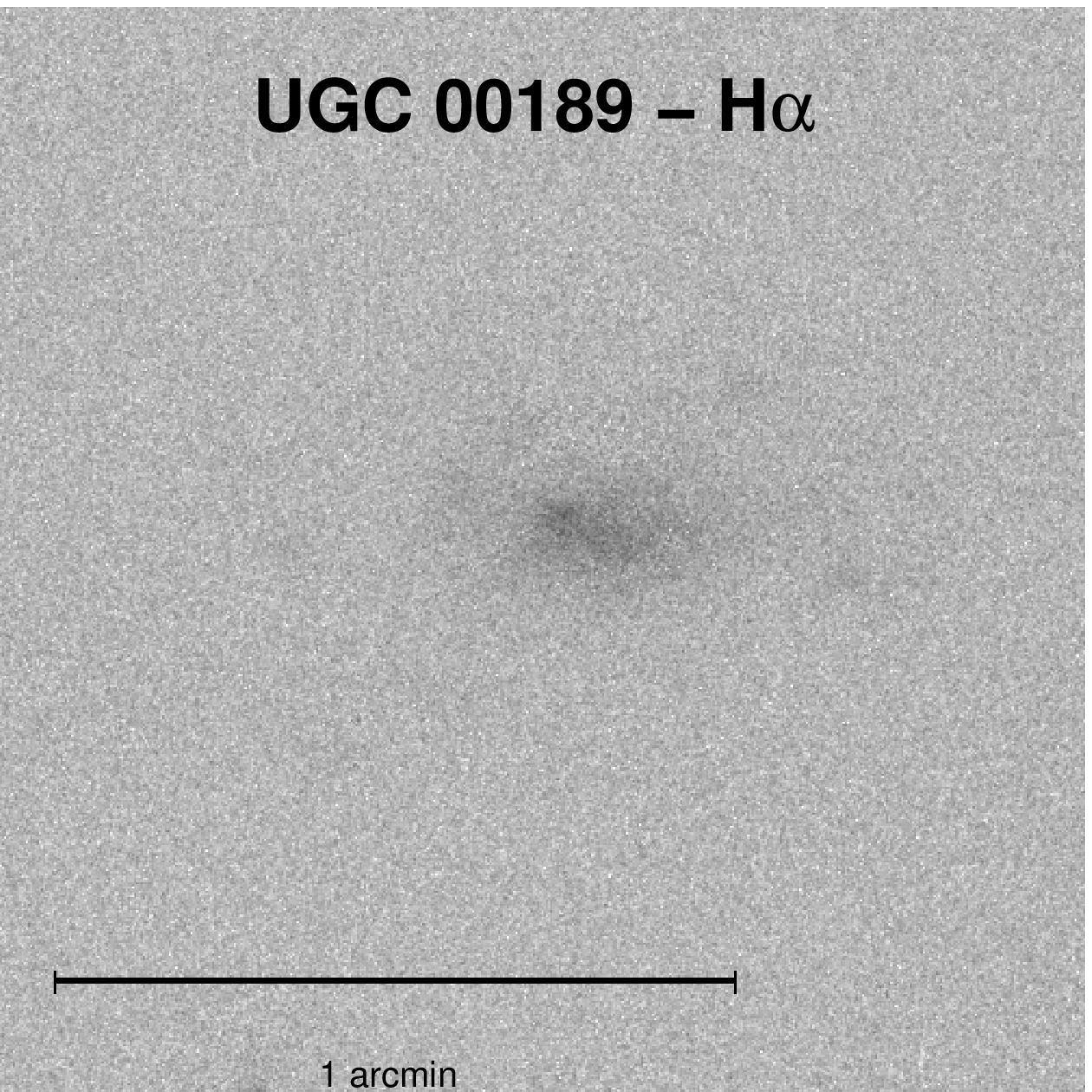} }
\resizebox{4.4cm}{!}{\includegraphics{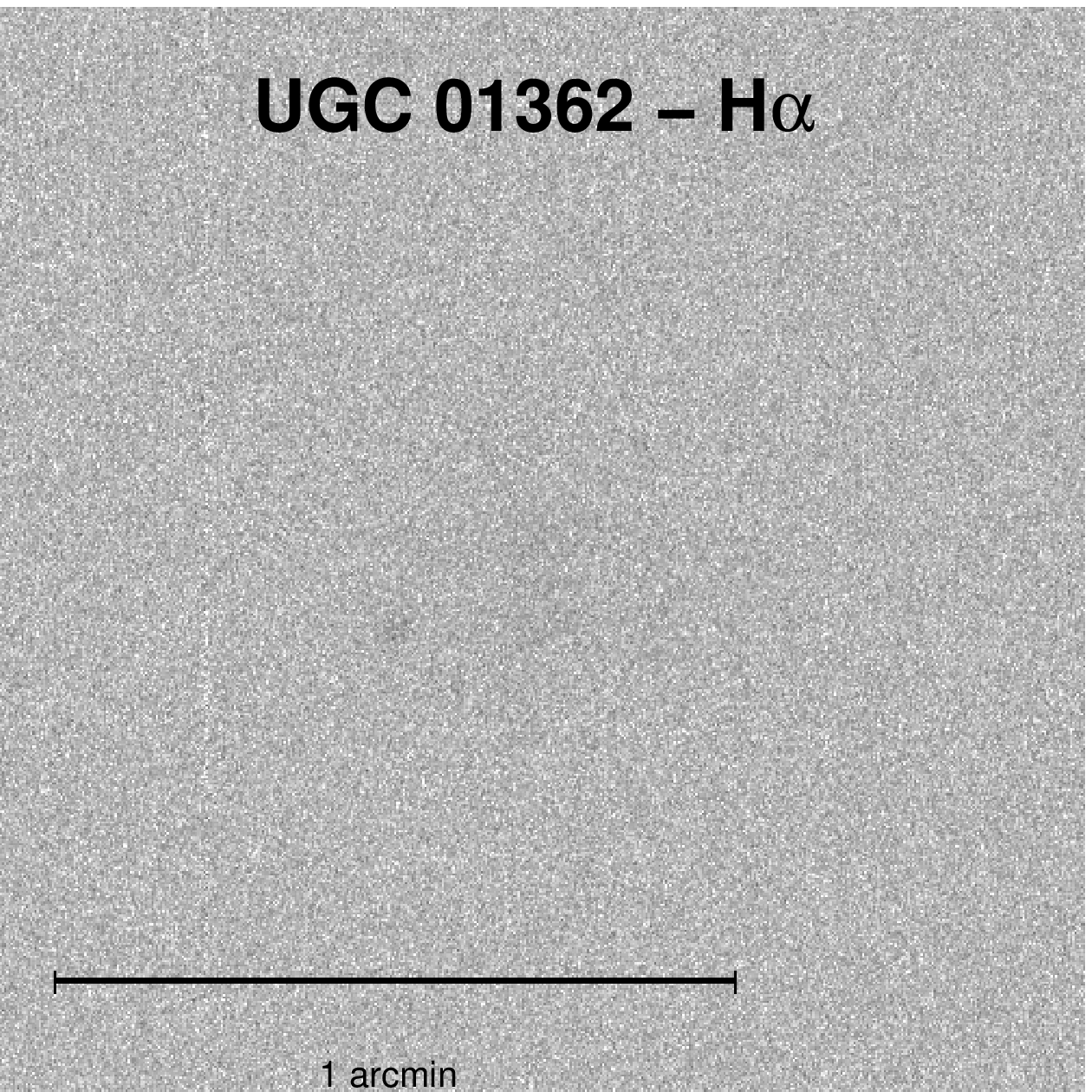} } }
\caption[]{H$\alpha$ images of three low surface brightness galaxies, showing the diversity in 
morphologies found by \cite{oneil07}.  
}\label{fig:Halpha_images} 
\end{figure}

\subsection{Star Formation -- H$\alpha$ and UV Images}
A recent paper by \cite{oneil07} shows the H$\alpha$ morphology of massive LSB galaxies to be more varied than the galaxies' optical distribution.  The images obtained show H$\alpha$ morphologies ranging from no H$\alpha$ detection through galaxies with well defined arms and a bright nuclear bulge and including galaxies with H$\alpha$ visible only in star forming clumps in the galaxies' outer arms (Figure~\ref{fig:Halpha_images}).  

\begin{figure}
\centering
\resizebox{6.5cm}{!}{\includegraphics{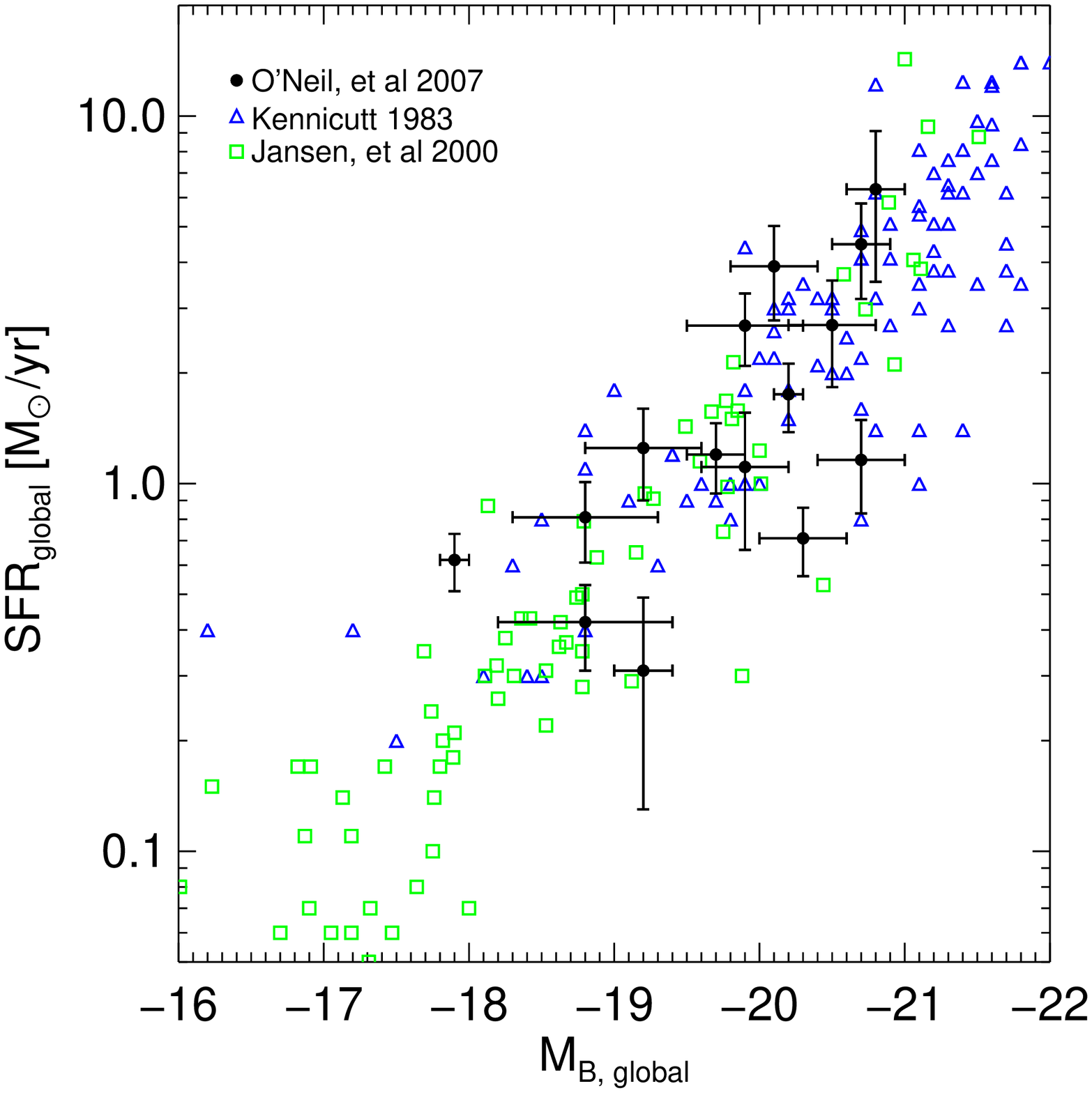} }
\resizebox{6.5cm}{!}{\includegraphics{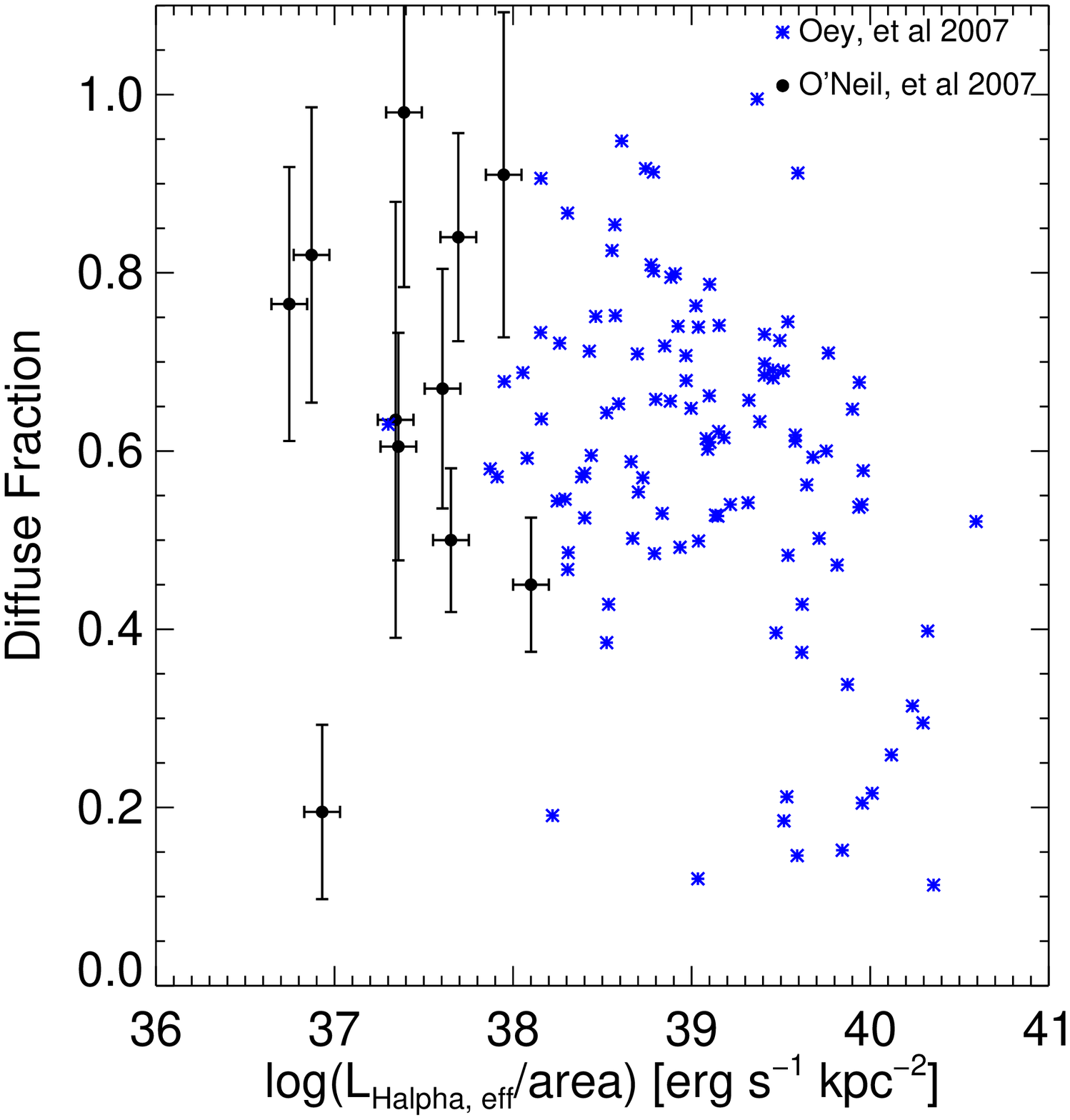} }
\caption[]{\textit{Right:} Total B magnitude plotted against star formation rate. \textit{Left:} Luminosity surface brightness
(Luminosity/area) plotted against the diffuse H$\alpha$ fraction for \cite{oneil07} and that of \cite{oey07}. 
In both plots the lack dots indicate the massive lower surface brightness galaxies in \cite{oneil07}.   
}\label{fig:Halpha}
\end{figure}

\cite{oneil07} also look at the global star formation rate, as determined by the H$\alpha$ emission, and at the distribution of the H$\alpha$ gas within the studied galaxies.  Here they find two somewhat surprising facts, as shown in Figure~\ref{fig:Halpha}.  First, the global star formation rate of the lower surface brightness galaxies is the same as for high surface brightness galaxies, based off the galaxies' total (B) luminosity.  Second, the diffuse fraction of the lower surface brightness galaxies, defined by the total amount of H$\alpha$ gas {\it not} contained within HII regions compared with the total H$\alpha$ found within the galaxies, is much higher for the lower surface brightness sample than the comparison high surface brightness sample.  Combined, these two facts indicate that considerable star formation is occurring outside the HII regions, similar to the findings of \cite{thilker05} for the outer (low surface brightness) edges of M83.

The best method for determining the total star formation rate for the massive LSB galaxies is through obtaining complementary UV images of the galaxies, similar to the study done by \cite{thilker05} for M83.  GALEX NUV and FUV images of 8 massive LSB galaxies have been obtained and are currently being reduced.  Results from that study should be published near the end of 2007 and will allow for a more detailed look into the recent star formation within these intriguing systems.

\subsection{Dust}
\begin{figure}
\centering
\resizebox{6.5cm}{!}{\includegraphics{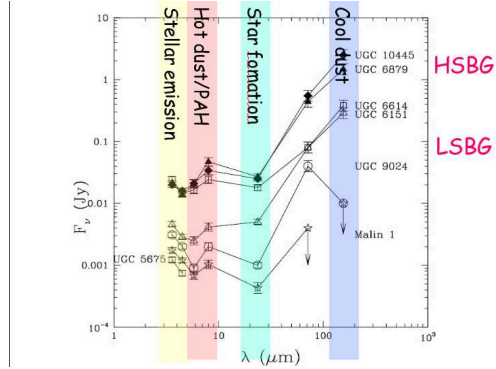} }
\caption[]{Spectral energy distribution of all the galaxies in \cite{hinz07}.  The arrows represent 3$\sigma$
upper limits at 70 and 160 $\mu$m.  This is a modified version of Figure 6 in \cite{hinz07}.
\label{fig:spitzer} }
\end{figure}

The dust in massive LSB galaxies can be studied through directly measuring the hot and cool dust emission in the infrared.  \cite{hinz07} recently published the results of a SPITZER IR survey of four massive LSB galaxies, two of which were not detected above 12$\mu$m and two of which were (Figure~\ref{fig:spitzer}).  The two galaxies detected at 160$\mu$m provide the first direct measurements of cool dust within any LSB galaxy. \cite{hinz07} conclude that the strength of dust emission is dependent on the existence of bright star forming regions within the galaxies, but with only two detections in the 160$\mu$m band no general conclusions can be drawn from the data.

\subsection{CO Content}
\begin{figure}
\centering
\resizebox{10.5cm}{!}{\includegraphics{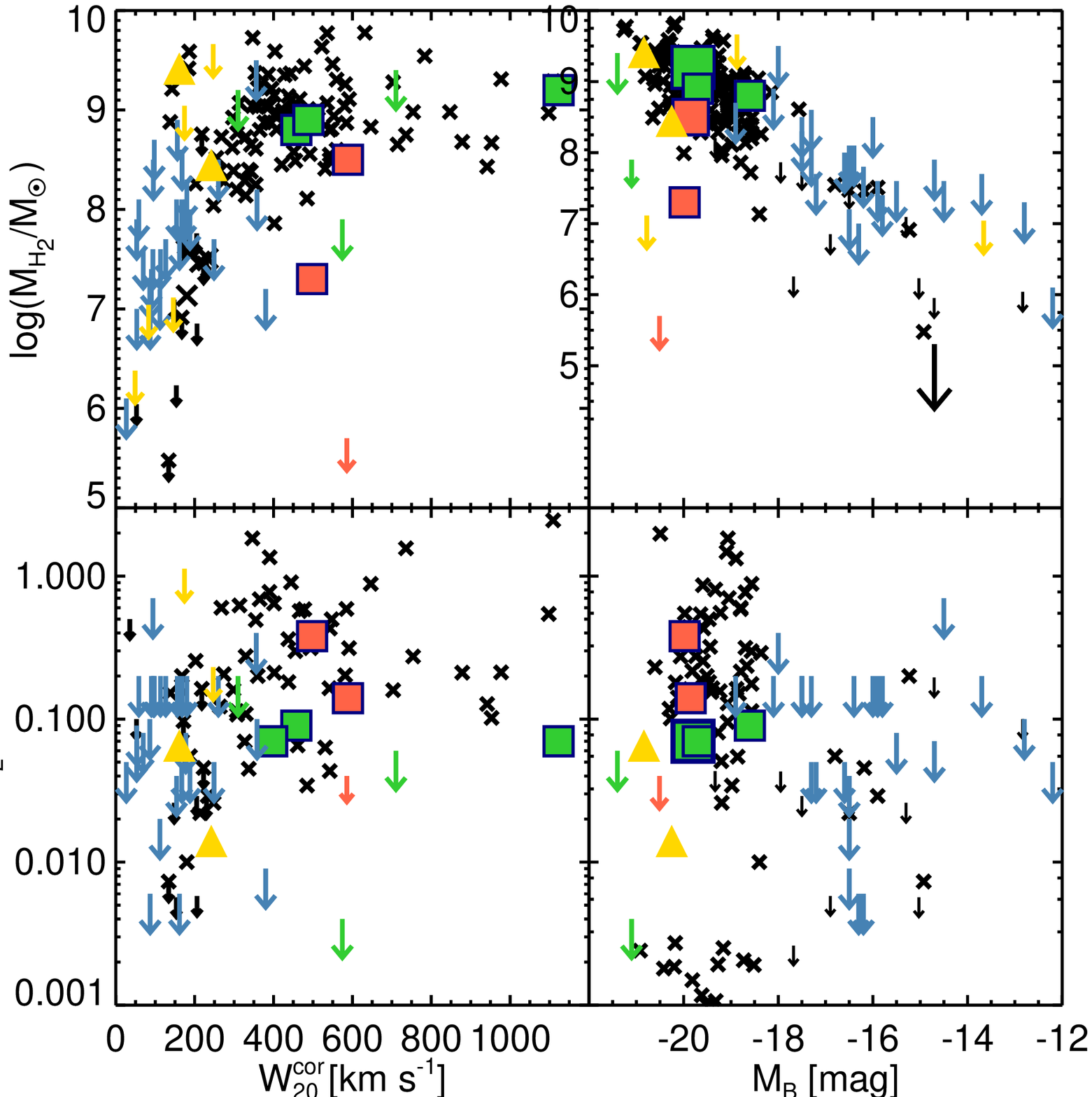} }
\caption[]{Inclination corrected HI velocity widths versus
H$_2$ mass (top left) and the H$_2$-to-HI mass
ratio (bottom left).  At right is the absolute B magnitude versus H$_2$ mass
(top right) and the H$_2$-to-HI mass
ratio (bottom right).  The squares (red \& green) are LSB galaxies from \cite{oneil04},
\cite{oneil03}, and \cite{oneil00}).
The (gold) triangle are LSB galaxies from \cite{das06}, 
the (blue) arrows are previous LSB (lower mass) measurements (\cite[O'Neil, Hofner, \& Schinnerer 2000]{oneil00};
\cite[Braine, Herpin, \& Radford 2000]{braine00}; \cite[de Blok \& van der Hulst 1998]{deblok98};
\cite[Schombert, \etal\ 1990]{schombert90})
and the (black) crosses are taken from various studies of the CO content in
HSB spiral galaxies (\cite[Casoli, \etal\ 1996]{casoli96}; \cite[Boselli, \etal\ 1996]{boselli96};
\cite[Tacconi \& Young 1987]{tacconi87}) and from the \cite{matthews01}
study of extremely late-type, edge-on spiral galaxies.
An arrow indicates only an upper limit was found.  Note that only 5 massive LSB galaxies
are shown in the plots on the right, as the absolute magnitude of LSBC F582-2 is
not known.  
}\label{fig:CO}
\end{figure}

The total number of massive LSB galaxies with CO measurements is small -- to date only 
10 massive LSB galaxies have even been surveyed for CO (\cite[Das, \etal\ 2006]{das06}; \cite[O'Neil \& Schinnerer 2004]{oneil04};
\cite[O'Neil, Schinnerer, \& Hofner 2003]{oneil03}; \cite[O'Neil, Hofner, \& Schinnerer 2004]{oneil04}). 
Yet compared to less massive LSB galaxies, the number of massive LSB galaxies with CO
detections is high, as 60\% of the studied galaxies have had some CO detected, as compared to a $<$3\% detection rate for less massive LSB galaxies (Figure~\ref{fig:CO}). 
Of the two CO maps published to date, one shows the CO to lie only within the galaxy's bright nuclear
bulge (\cite[O'Neil \& Schinnerer 2003]{oneil03}) while the other shows a clear CO detection within
the galaxy's low surface brightness disk (\cite[Das \etal\ 2006]{das06}).  A number of additional maps
of CO within LSB galaxies have recently been obtained and should be published soon (\cite[O'Neil, \etal\ 2008a]{oneil08}).

\subsection{Radio Continuum \& AGN}
Only a few studies have been done to date examining the radio continuum properties of LSB galaxies. 
\cite{oneil06} used archival data to measure the galaxies' FIR/1.4 GHz ratio and found that at least
one of the 28 galaxies studied appears to harbor an AGN. \cite{das07} has undertaken a more thorough study of a
few massive LSB galaxies and found one of the three galaxies studied to date has clear AGN indicators.

\section{Putting it all Together -- A Multi-wavelength Study of Massive LSB Galaxies}

While fascinating on their own, the individual studies described above become much more interesting when gathered together to present a multi-wavelength view of massive LSB galaxies.  This is the topic of considerable ongoing research by the author and numerous collaborators.  Over the next year we should be completing a number of studies, including:
\begin{itemize}
\item An HI survey of more than 300 massive potentially LSB galaxies with unknown redshifts using the Arecibo, Nan\c{c}ay and Green Bank single dish radio telescopes (\cite[O'Neil, \etal\ 2008a]{oneil08});
\item Six detailed HI maps of massive LSB galaxies obtained with the VLA and GMRT (\cite[O'Neil, Harnett, \& Schinnerer 2008]{oneil08b});
\item Three additional CO maps of massive LSB galaxies previously detected with the IRAM 30m telescope (\cite[O'Neil, van Driel, \& Combes 2008]{oneil08c});
\item GALEX NUV and FUV images of five massive LSB galaxies (\cite[O'Neil, Baker, \& Mitchell 2008]{oneil08d});
\end{itemize}
In addition to the above, our group plans on obtaining Spitzer images and spectra, detailed HI maps,
higher resolution optical images, and high resolution rotation curves of a larger sample of massive LSB galaxies.
When complete, these studies will be combined to provide the most complete view onto the star formation history and potential of massive LSB galaxies done to date.

\begin{acknowledgments}
This work could not be done without the help of all my collaborators -- A. Baker, G. Bothun, F. Combes, J. Harnett, S. Oey, S. Schneider, E.  Schinnerer, W. van Driel.  And of course without the help of Paul, Max, and Willie.
\end{acknowledgments}

\end{document}